\RequirePackage{lineno}
\documentclass[a4paper,10pt,twoside,preprint,aps]{cpc-hepnp}
\usepackage{CJK,upgreek,fancyhdr}
\usepackage{multicol}
\usepackage{graphicx}
\usepackage{booktabs}
\usepackage{amssymb,bm,mathrsfs,bbm,amscd}
\usepackage[tbtags]{amsmath}
\usepackage{lastpage}
\usepackage{epsfig}
\usepackage{epsf}
\usepackage{latexsym}
\usepackage{dcolumn}
\usepackage{bm}
\usepackage{subfigure}
\usepackage{comment}
\def \ee {\mathrm{e^+e^-}}
\def \pp {\mathrm{\uppi^+\uppi^-}}
\def \kk {\mathrm{K^+K^-}}
\def \jp {\mathrm{J/\uppsi}}
\def \psip {\mathrm{\uppsi(2S)}}

\def \lunda {{\small\rm LUARLW~}}
\def \mm {\mathrm{\upmu^+\upmu^-}}
\def \gg {\mathrm{\upgamma\upgamma}}
\def \logo {Where the points with errors are data, and shaded histogram is MC distribution.}

\begin{document}
\begin{CJK*}{GB}{gbsn}
\fancyhead[c]{\small Chinese Physics C~~~Vol. xx, No. x (201x) xxxxxx}
\fancyfoot[C]{\small 010201-\thepage}
\footnotetext[0]{Received 31 June 2015}

\title{Tuning and validation of hadronic event generator for $R$ value measurements in the tau-charm region
\thanks{Supported by National Natural Science
Foundation of China, under Contract NO. 11175146, 11375205, 11575077, 11335008, 11565006, Large Science Setup of Joint Foundation 10979059, and 100 Talents Program of CAS}
}
\author{
Rong-Gang Ping (平荣刚)$^1$\email{pingrg@ihep.ac.cn, zhengbo\_usc@163.com} Xi-An Xiong (熊习安)$^1$ Lei Xia (夏磊)$^2$ Zhen Gao (高榛)$^2$\\
Ying-Tian Li (李应天)$^1$ Xing-Yu Zhou (周兴玉)$^1$ Bing-Xin Zhang (张丙新)$^1$ Bo Zheng (郑波)$^3$ \\
Wen-Biao Yan (鄢文标)$^2$ Hai-Ming Hu (胡海明)$^1$ Guang-Shun Huang (黄光顺)$^2$ }

\maketitle

\address{%
$^1$Institute of High Energy Physics,Chinese Academy of Sciences, Beijing 100049, People's Republic of China\\
$^2$University of Science and Technology of China, Hefei 230026, People's Republic of China\\
$^3$University of South China, Hengyang, 421001, People's Republic of China
}

\large

\begin{abstract}
To measure the $R$ value in an energy scan experiment with $\ee$ collisions, precise calculation of initial state radiation is required in the event generators. We present an event generator for this consideration, which incorporates initial state radiation effects up to second order accuracy. The radiative correction factor is calculated using the totally hadronic Born cross section. The measured exclusive processes are generated according to their cross sections, while the unknown processes are generated using the LUND Area Law model, and its parameters are tuned with data collected at $\sqrt s=3.65$ GeV. The optimized values are validated with data in the range $\sqrt s=2.2324\sim3.671$ GeV. These optimized parameters are universally valid for event generation below the $\mathrm{D\bar D}$ threshold.
\end{abstract}

\begin{keyword}
Event generator, R value
\end{keyword}

\begin{pacs}
13.66.Jn, 02.70.Uu
\end{pacs}


\section{Introduction}
The total cross section for hadron production in positron-electron ($\ee$) annihilation is one of the most fundamental observables in particle physics. A precise measurement of the hadronic cross section allows us to
determine the hadronic contributions to the running of the quantum electrodynamic (QED) fine structure constant $\alpha$, electroweak parameters, and the strong coupling $\alpha_s$. The $R$ value, defined as the ratio of the total hadronic cross section to that of  $\ee\to\mm$ at Born level, have been measured by many collaborations in $\ee$ scan experiments, over the center-of-mass energy from the two pion mass threshold ($M_{2\uppi}$) to the $\mathrm{Z}$ peak \cite{Rvalues}. In the tau-charm energy region, the $R$ values measured at BESII \cite{bes2R} were used in the evaluation of the hadronic contribution from
the five quark loops at the energy of $\mathrm{Z}$ peak, $\Delta\alpha_{\textrm{had}}^{(5)}(M_\mathrm{Z}^2)$, with an improved precision by a factor of 2 \cite{alpha}.

A large number of exclusive processes have been measured over the range from $M_{2\uppi}$ to 5 GeV \cite{xscollection}, but most cross sections have large uncertainties. To improve these measurements, a hadronic event generator is needed for us to get better understanding of background events from $\ee\to$ hadrons.

Especially, a precise $R$-value measurement requires excellent control of radiative correction (RC) and vacuum polarization (VP) in the Monte Carlo (MC) program. We design an event generator for measuring $R$ values and exclusive decays in $\ee$ collisions. The generator is constructed in the framework of BesEvtGen \cite{besevtgen}, incorporating both the RC and VP effects. We also present details of the parameter optimization of the Lund Area Law (\lunda) model \cite{lundamodel} with data, and validations with various distributions within the energy range $\sqrt s=2.2324\sim3.671$ GeV.

\section{Framework of event generator}
The generator is constructed as a model of the BesEvtGen package. It provides the 4-momentum of each final state particle for detector simulation, and provides the ISR correction factor and VP factors for users to undress the observed cross section. The basic idea of this generator is to decompose the total hadronic cross section into the measured exclusive processes and remaining unknown processes. The latter are generated with the \lunda model.
\subsection{Initial state radiative correction}
\begin{figure}[htbp]
\begin{center}
\epsfysize=4cm \epsffile{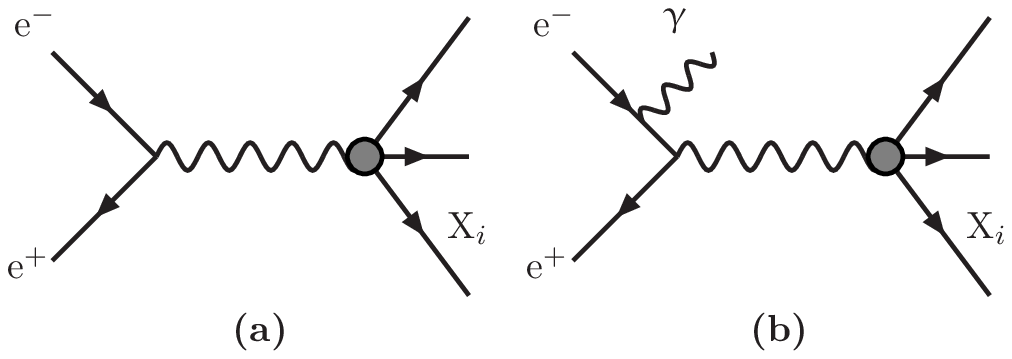}
\caption{Feynman diagrams for the process (a) $\ee\to \mathrm{X_i}$, and ISR process (b) $\ee\to\mathrm{\upgamma}_{\rm ISR} \mathrm{X_i}$  \label{fig1}.}
\end{center}
\end{figure}

In an $\ee$ energy scan experiment, we consider a measurement of the Born cross section ($\sigma_0$) for a process
$\ee\to \mathrm{X_i}$, as shown in Fig. \ref{fig1} (a), where $\mathrm{X_i}$ denotes the hadron states of $\mathrm{i}$-th process. Due to ISR, the observed cross section ($\sigma$) is actually for the process $\ee\to\upgamma_{\rm ISR}\mathrm{X_i}$, as shown in Fig. \ref{fig1} (b). The observed cross section is related to the Born cross section by the quasi-real electron method \cite{bornXS}:
\begin{equation}\label{master}
\sigma(s)=\int_{M_\textrm{th}}^{\sqrt s} dm {2m\over s}W(s,x){\sigma_0(m)\over |1-\Pi(m)|^2},
\end{equation}
where $m$ is the invariant mass of the final states; $\Pi(m)$ is the vacuum polarization function, which will be discussed later; $s$ is the $\ee$ center-of-mass energy squared;  $x\equiv2E^*_\upgamma/\sqrt s=1-m^2/s$, and $E^*_\upgamma$ is the total energy carried by ISR photons in the $\ee$ center-of-mass frame; $M_{\rm{th}}$ is the mass threshold of a given process.

To calculate the finite-order leading logarithmic correction, the structure function method is used \cite{fadin}. This method results in the same factorized form for the radiative photon emission cross section. Up to order $\alpha^2$, the radiative function takes the form:
\begin{equation}{\label{secondRad}}
W(s,x)=\Delta\beta x^{\beta-1}-{\beta\over 2}(2-x) + {\beta^2\over 8}\{(2-x)[3\ln(1-x)-4\ln x]-4{\ln(1-x)\over x}-6+x\},
\end{equation}
where
\begin{eqnarray}
L&=&2\ln {\sqrt s \over m_\mathrm{e}},\\
\Delta&=&1+{\alpha\over\uppi}({3\over 2}L+{1\over 3}\uppi^2-2)+({\alpha\over \uppi})^2\delta_2,\\
\delta_2&=&({9\over 8}-2\xi_2)L^2-({45\over 16}-{11\over 2}\xi_2-3\xi_3)L-{6\over 5}\xi_2^2-{9\over 2}\xi_3-6\xi_2\ln 2+{3\over 8}\xi_2+{57\over 12},\nonumber\\
\beta&=&{2\alpha\over \uppi}(L-1),~\xi_2=1.64493407,~\xi_3=1.2020569.
\end{eqnarray}
Here the exponential part in Eq. (\ref{secondRad}) accounts for soft multi-photon
emission, while the remaining part takes into account hard collinear bremsstrahlung
in the leading logarithmic approximation. We use the radiative function up to the second
order calculation to determine the cross section; it is accurate enough to construct the event generator for our purpose, though contributions from the $\alpha^3$-order are known \cite{thirdRad}.

\begin{figure}
\centering
\begin{tabular}{cc}
  \epsfig{file=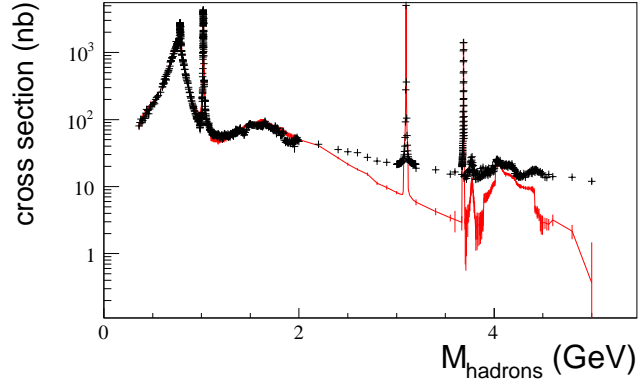,width=0.5\linewidth,clip=}
  \end{tabular}
  \caption{(color online) The cross section for light hadron production within $M_{2\uppi}\sim$5 GeV, where the black points with errors are the total hadronic cross section \cite{pdg}, and the histogram with points (in red) is the sum of measured cross sections for exclusive processes.
  }
  \label{xssum}
\end{figure}

To do the RC for the process $\ee\to$ hadrons, we use the cross sections for the light hadron productions measured so far.
In the energy region from $M_{2\uppi}$ to 5 GeV, the total cross sections are quoted from the Particle Data Group (PDG) \cite{pdg}. The total distribution is shown in Fig. \ref{xssum}.

At the leading order of QED calculation, the ISR photon is characterized by soft energy and beam collinear distribution. A more general result is obtained by the method of Bonneau and Martin\cite{bonneau} up to $m^2_\mathrm{e}/s$ terms, and the angular distributions is calculated by
\begin{eqnarray}\label{angISR}
{d\sigma(s,x)\over dx d\cos\theta}&=&{2\alpha\over \uppi x}(1-x+{x^2\over 2})\sigma_0(s(1-x)) P(\theta), \textrm{~with~}\\
P(\theta)&=&\frac
{\sin^2{\theta}-\frac{x^2\sin^4{\theta}}{2(x^2-2x+2)}-
\frac{m_\mathrm{e}^2}{E^2}~\frac{(1-2x)\sin^2{\theta}-x^2\cos^4{\theta}}{x^2-2x+2}}
{\left ( \sin^2{\theta}+\frac{m_\mathrm{e}^2}{E^2}\cos^2{\theta}
\right )^2 },
\end{eqnarray}
where $E$ is the beam energy in the center of mass system of the electron
and positron.
\subsection{Vacuum polarization}

The VP of the photon is a quantum effect which leads to the scale dependence of the electromagnetic coupling. It therefore plays an important role in the $\ee$ physics and it is crucial to know it for the $R$ value measurement.

Conventionally the VP function is denoted by $\Pi(q^2)$, where $q$ is a space- or time-like momentum. In the $R$ value measurement, we only consider the time-like case, i.e. $s=q^2$, which receives all possible one-particle irreducible leptonic and hadronic contributions. Their contributions to $\Pi(s)$ are calculated and then summed. While the leptonic contributions can be predicted within perturbative theory, the precise determination of the hadronic contributions depends on dispersion relations using experimental data as input.

\begin{figure}
 \centering
  \epsfig{file=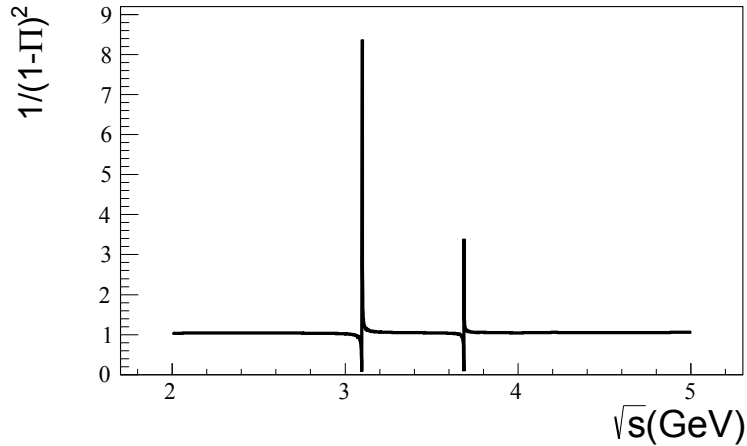,height=6cm}
  \caption{Vacuum polarization factor $1/|1-\Pi(s)|^2$ quoted from Ref. \cite{vp} \label{vpfj}.}
  \label{vp}
\end{figure}

The VP has been calculated by many groups and is available in the literature. Comparisons between them are given in Ref. \cite{actis}. There are notable differences below 1.6 GeV, and above 2.0 GeV; visible differences appear when approaching the charmonium resonances. We use the results from the Fred Jegerlehner group \cite{vp}. It provides leptonic and hadronic
VPs both in the space- and time-like region. For the leptonic
VP the complete one- and two-loop results and the
known high-energy approximation for the three-loop corrections
are included. The hadronic contributions are given
in tabulated form in the subroutine HADR5N \cite{hadr5n}. Figure \ref{vpfj} shows the
VP factor defined by $1/|1-\Pi(s)|^2$ in the energy region $\sqrt s=2.0-5$ GeV. The values at $\jp$ and $\psip$ peaks are very large but less significant elsewhere.

\subsection{Cross sections for exclusive processes}
Many exclusive processes have been measured in the $\ee$ collision experiments. Currently we collect 76 exclusive modes, with energy region covering from 0.3 GeV up to about 6 GeV. The Born cross sections are quoted from the published papers; their information is given in Table \ref{exclusiveXS}. The sum of these cross sections is shown in Fig. \ref{xssum}. Below 2.0 GeV, the total cross section is the sum of the exclusively measured ones.

\begin{table}
\caption{Collection of measured exclusive processes. Their cross sections are quoted from the references as given in table, together with the energy ranges.\label{exclusiveXS}}
\begin{center}
\begin{tabular}{clll||clll}
\hline\hline
ID & $\ee\to$    & $\sqrt{s}$ (GeV) & Reference&ID & $\ee\to$    & $\sqrt{s}$ (GeV) & Reference\\\hline
1  & $\mathrm{p\bar p}$   & 1.877-4.500    &\cite{babar_ppbar}&  39 & $\upomega\pp $ &1.150 - 2.525 & \cite{babar_llbar}\\
2  & $\mathrm{n\bar n}$   & 1.90 - 2.44    &\cite{infn} &        40 & $\upomega \mathrm{f_0(980)} $ &1.700 - 2.475 &\cite{babar_llbar}\\
3  & $\Lambda\bar \Lambda$ &2.23 - 5.00 &\cite{babar_llbar}&    41 & $\upeta'\pp $ &1.58 - 3.42 &\cite{babar_llbar}\\
4  & $\Sigma\bar \Sigma^0$ &2.385 - 5.000&\cite{babar_llbar}&   42 & $\mathrm{f_1(1285)}\pp $ &1.66 - 3.50 & \cite{babar_llbar}\\
5  & $\Lambda\bar \Sigma^0$ &2.308 - 5.000 & \cite{babar_llbar}& 43 & $\upomega\kk $ &1.57 - 3.45  &\cite{babar_llbar}\\
6  & $\Sigma^0\bar \Lambda $&2.308 - 5.000 &\cite{babar_llbar}&  44 & $\upomega\pp\uppi^0 $ &1.500 - 4.423  &\cite{babar_4pi}\\
7  & $\uppi^+\uppi^- $ &0.305 - 2.950 &\cite{babar_pipi}&   45 & $\Sigma^-\bar \Sigma^+ $ &2.308 - 5.000 &\cite{babar_llbar}\\
8  & $\uppi^+\uppi^-\uppi^0 $ &1.063 - 2.989 &\cite{barbar_3pi}&    46 & $\mathrm{K^+K^-}$ &1.009 - 4.170 & \cite{kkbar,ppkkpipi}\\
9  & $\mathrm{K^+K^-}\uppi^0 $ &1.34 - 4.68 & \cite{babar_kkpi0}&  47 & $\mathrm{K_{S}K_{L}}$ &1.004 - 2.140 & \cite{kkbar}\\
10 & $\mathrm{K_SK^+}\uppi^- $ &1.26 - 4.66&\cite{babar_kkpi0}&    48 & $\upomega\upeta$ &1.371 - 3.178 & \cite{babar_6pi} \\
11 & $\mathrm{K_SK^-}\uppi^+$ &1.26 - 4.66 & \cite{babar_kkpi0}&   49 & $\mathrm{p\bar p}\uppi^0$ &4.009 - 4.200 &\cite{liujie} \\
12 & $\mathrm{K^+K^-}\upeta $ &1.69 - 3.13 &\cite{babar_kkpi0}&    50 & $\mathrm{p\bar p}\upeta$ &4.009 - 4.200  &\cite{liujie}\\
13 & $2(\pp) $    &0.615 - 4.45 &\cite{babar_4pi} &     51 & $\mathrm{D^-D^{*0}}\uppi^+$ &4.020 -5.171& \cite{ddstarpi}\\
14 & $\pp2\uppi^0$  &0.185 - 2.98 &\cite{babar_pipi2pi0}& 52 & $\mathrm{D^+D^{*0}}\uppi^-$ &4.020 -5.171& \cite{ddstarpi}\\
15 & $\kk\pp $ & 1.425 - 4.988 &\cite{babar_kkpp} &    53 & $\mathrm{D^{*0}\bar~D^{*0}}$ & 4.033 -4.991 & \cite{ddstar}\\
16 & $\kk 2\uppi^0 $ &1.50 - 4.02 & \cite{babar_kkpp}&     54 & $\mathrm{D^0\bar D^{*0}}$ &4.033 -4.991 &\cite{ddbar}\\
17 & $2(\kk) $ &2.02 -4.54 &\cite{babar_4k}&    55 & $\mathrm{\bar D^0 D^{*0}}$&3.814 - 4.990&\cite{ddbar}\\
18 & $2(\pp)\uppi^0$ &1.013- 4.488&\cite{babar_5pi}&    56 & $\mathrm{D^0\bar D^0}$ &3.814 - 4.990&\cite{ddbar}\\
19 & $2(\pp)\upeta $ &1.313 - 4.488&\cite{babar_5pi}&   57 & $\mathrm{D^+D^-}$ &3.814 -4.990 &\cite{ddbar}\\
20 & $\kk\pp\uppi^0 $ &1.613 - 4.488& \cite{babar_5pi}& 58 & $\mathrm{D^+D^{*-}}$ &3.890 -4.994& \cite{ddstar}\\
21 & $\kk\pp\upeta $  &2.113 - 4.488 &\cite{babar_5pi}& 59 & $\mathrm{D^-D^{*+}}$ &3.890 -4.994 &\cite{ddstar}\\
22 & $3(\pp) $   &1.313 - 4.488 & \cite{babar_6pi}&   60 & $\mathrm{D^{*+}D^{*-}}$ &4.033 -4.991& \cite{ddstar}\\
23 & $2(\pp\uppi^0)$ &1.313 - 4.488  & \cite{babar_6pi}&        61 & $\mathrm{D^0D^-}\uppi^+$&4.015 -4.974 &\cite{ddpi}\\
24 & $\upphi\upeta $  &1.57 - 3.45&\cite{babar_kkpi0}&      62 & $\mathrm{\bar D^0D^+}\uppi^-$ &4.015 -4.974&\cite{ddpi}\\
25 & $\upphi\uppi^0 $ &1.25 - 1.45 &\cite{babar_kkpi0}&     63 & $\mathrm{D^0D^{*-}}\uppi^+$&4.020 -5.171&\cite{ddstarpi}\\
26 & $\mathrm{K^+K^{*-}} $ &1.37 -1.99  &\cite{babar_kkpi0}&     64 & $\mathrm{\bar D^0D^{*+}}\uppi^-$ &4.020 -5.171&\cite{ddstarpi}\\
27 & $\mathrm{K^-K^{*+}} $ &1.37 - 1.99 &\cite{babar_kkpi0}&     65 & $\uppsi(2S)\uppi^0\uppi^0$&4.127-5.480& \cite{pipipsip}\\
28 & $\mathrm{K_S\bar K^{*0}(892)} $ &1.37 - 1.99 &\cite{babar_kkpi0}&   66 & $\upeta\jp$ &3.81 -4.68&\cite{etajsi}\\
29 & $\mathrm{K^{*}(892)^0K^+}\uppi^- $ &1.588 - 3.963 & \cite{babar_kkpp}&      67 & $\uppi^+\uppi^-\mathrm{h_c}$ &4.009-4.420&\cite{pippimhc}\\
30 & $\mathrm{K^{*}(892)^0K^-}\uppi^+$ &1.588 - 3.963 & \cite{babar_kkpp}&       68 & $\uppi^0\uppi^0\mathrm{h_c}$ &4.009-4.420&\cite{pizpizmhc}\\
31 & $\mathrm{K^{*}(892)^-K^+}\uppi^0 $ &1.588 - 3.963 &\cite{babar_kkpp}&       69 & $\mathrm{K^+K^-}\jp$ & 4.179 -4.970 &\cite{kkjsi}\\
32 & $\mathrm{K^{*}(892)^+K^-}\uppi^0 $ &1.588 - 3.963 & \cite{babar_kkpp}&      70 & $\mathrm{K_S^0K_S^0}\jp$ &4.179 -4.970&\cite{ksksjsi}\\
33 & $\mathrm{K_2^*(1430)^0K^+}\uppi^- $&2.348 - 3.965 &\cite{babar_kkpp}& 71 & $\jp\pp$ &3.829 -5.471&\cite{pipijsi}\\
34 & $\mathrm{K_2^*(1430)^0K^-}\uppi^+ $ &2.348- 3.965 &\cite{babar_kkpp}& 72 & $\psip\pp$ &4.127 -5.480&\cite{pipipsip}\\
35 & $\kk\uprho $ &1.777 - 3.830 &\cite{babar_kkpp}&       73 & $\mathrm{D_s^+D_s^-}$ &3.97 -4.26 &\cite{dsds}\\
36 & $\upphi\pp $ &1.488 - 2.863&\cite{babar_kkpp}&     74 & $\mathrm{D_s^{*+}D_s^-}$ &4.12 -4.26&\cite{dsds}\\
37 & $\upphi \mathrm{f_0(980)} $ &1.888- 2.963 & \cite{babar_kkpp}&      75 & $\mathrm{D_s^{*-}D_s^+}$ &4.12 -4.26 &\cite{dsds}\\
38 & $\upeta\pp $ &1.025 - 2.975 & \cite{babar_llbar}&   76 & $\Lambda_c^+\Lambda_c^-$&4.57 -4.64&\cite{lambdacpair}\\\hline\hline
\end{tabular}
\end{center}
\end{table}

The narrow vector resonances, such as $\uppsi(3770)$, $~\psip$, $~\jp$, $~\uprho(1700)$, and $~\upomega(1420)$, are also included in the calculation for the ISR correction factor. The cross sections for these narrow resonances are represented with the Breit-Wigner function
 $$\sigma_{BW}(s)=12\uppi{\gamma_{\mathrm{ee}}\gamma\over (s-M^2)+M^2\gamma^2}, $$
where $M,~\gamma,$ and $\gamma_{\mathrm{ee}}$ are the mass, total width and partial decay width to $\ee$ final state, respectively.

The distribution of cross section versus center-of-mass energy is described by an empirical function, which is parameterized with a multi-Gaussian function. Its parameters are determined by fitting the cross section mode by mode. These empirical functions are used in the generator for the calculation of the ISR correction factor and event type sampling.

The angular distribution for ISR photons is implemented according to Eq. (\ref{angISR}). However, angular distributions are implemented only for two-body decays, namely, $1-\cos^2\theta$ for $\mathrm{PP}$ (where $\mathrm{P}$ is a pseudoscalar meson) modes, and $1+\alpha\cos^2\theta$ for the $\mathrm{PV}$ ($\alpha=1$) and $\mathrm{B\bar B}$ modes, where $\mathrm{V}$ is a vector meson, and $\mathrm{B}$ is a baryon. The angular distribution parameter $\alpha$ for the $\mathrm{B\bar B}$ mode is taken as the quark model prediction \cite{bbbarang}. The phase space model is used for multi-body decays.

\subsection{LUND Area Law model}
The hadronic events produced in the $\ee$ annihilation are evolved as follows. As the first step, a quark-antiquark ($\mathrm{q\bar q}$) pair is produced from a virtual photon, coupled to the $\ee$ pair. Then the $\mathrm{q\bar q}$ branching proceeds via emitting gluons, and further develops into hadrons. In the high energy region, the cluster model (e.g. {\small\sc HERWIG} \cite{herwig}) and LUND string model (e.g. {\small\sc JETSET/PYTHIA}\cite{pythia6.4}) are available and precise enough to describe the hadronic fragmentation with parameters optimized at boson $\mathrm{Z}$ peak. However, in the intermediate and low energy region, parameters need to be optimized or a new model is desirable to describe the light quark fragmentation.

In the tau-charm energy region, the \lunda model \cite{lundamodel} has been proposed to estimate the multiplicity distribution for primary hadrons produced from the string fragmentation. The probability distribution reads:
\begin{equation}
P_n={\mu^n\over n!}\exp[c_0+c_1(n-\mu)+c_2(n-\mu)^2],
\end{equation}
with $\mu=\alpha+\beta\exp(\gamma\sqrt s)$, where $c_0,c_1,c_2,\alpha,\beta$ and $\gamma$ are parameters to be tuned with data. An interface to access the \lunda model is designed in the BesEvtGen \cite{besevtgen} framework, and is only used to generate the primary hadrons. The further decays into light hadrons are realized with BesEvtGen \cite{besevtgen}.

\subsection{Monte Carlo algorithm}
The event sampling proceeds via two steps. Firstly, the mass of the hadron system, $M_\textrm{hadrons}$, is sampled according to the distribution of the observed cross section, i.e. $d\sigma(s)/ dm$, for the process ${\ee\to\upgamma_{ISR} \mathrm{X_i}}$ according to Eq. (\ref{master}). For simplicity, the ISR energy, $\sqrt s-M_\textrm{hadrons}$, is imposed on a single photon. The second step is to sample the event type topology according to the ratios of individual cross sections at the energy point $M_\textrm{hadrons}$.

\subsubsection{Sampling of $M_\textrm{hadrons}$}
To calculate the total observed cross section at $\sqrt s$, we split the integral of Eq. (\ref{master}) into two parts, i.e.
\begin{equation}
\sigma(s)\equiv\sigma^I(s)+\sigma^{II}(s)=\int_{M_{\rm th}}^{M_0} dm {2m\over s}W(s,x){\sigma_0(m)\over |1-\Pi(m)|^2}
+\int_{M_{0}}^{\sqrt s} dm {2m\over s}W(s,x){\sigma_0(m)\over |1-\Pi(m)|^2},
\end{equation}
where the threshold energy $M_{\rm th}$ is the sum of masses for the final state particles,
and the broken point is taken at $M_0=\sqrt{s-2\sqrt sE_\upgamma^{\rm cut}}$ with a cut $E^{\rm cut}_\upgamma$ on the ISR photon energy. At the BESIII detector, the designed photon energy of the detection range is from $\sim20$ MeV to 4.2 GeV \cite{besIIINIM}. If the photon energy is less than 20 MeV, it will be impossible to reconstruct it in the detection simulation. Hence in practice, $E_\upgamma^{\rm cut}$ can be set to an energy less than the sensitivity of photon detection. e.g., $E_\upgamma^{\rm cut}=1$ MeV. In the range $0\sim E_\upgamma^{\rm cut}$, the ISR photon is too soft to be detected, so the ISR photon is not considered. To simplify the calculation, Born cross sections near the energy point $\sqrt s$ are assumed to be a constant value $\sigma_0(\sqrt s)$. Using the relation $x=1-m^2/s$, the second integral can be further decomposed into two parts:
\begin{equation}
\sigma^{II}(s)=\int_{M_{0}}^{\sqrt {s(1-b)}} dm {2m\over s}W(s,x){\sigma_0(m)\over |1-\Pi(m)|^2}+{\sigma_0(\sqrt s)\over |1-\Pi(\sqrt s)|^2}\lim_{a\to0}\int_a^bW(s,x)dx,
\end{equation}
with $b\ll1$. Using the radiative function given in Eq. (\ref{secondRad}), one has
\begin{eqnarray}
\lim_{a\to0}\int_a^bW(s,x)dx&=&\Delta b^\beta+{\beta^2b^2\over 32}+{\beta b^2\over 4}-{3\over16}\beta^2b^2\ln (1-b)
+{1\over 4}\beta^2 b^2\ln b-{5\over 16}\beta^2 b
-\beta b\nonumber\\
&&+{3\over 4}\beta^2b\ln(1-b)-\beta^2b\ln b-{9\over 16}\beta^2\ln(1-b)+{1\over 2}\beta^2\textrm{Li}_2(b),
\end{eqnarray}
with Spence's function $\textrm{Li}_2(x)=-x+{1\over 4}x^2-{1\over 9}x^3~(x\ll1)$.

To sample the $M_\textrm{hadrons}$, we split the region $M_\textrm{th}\sim \sqrt s$ into a few hundred intervals. The cumulative cross section up to the $i$-th interval, $m_i$, is $$\hat\sigma(m_i)={1\over \sigma(s)}\int_{M_\textrm{th}}^{m_i}dm {2m\over s}W(s,x){\sigma_0(m)\over |1-\Pi(m)|^2}.$$ The $M_\textrm{hadrons}$ is sampled according to the $\hat\sigma(m_i)$ distribution with the discrete MC sampling technique.

\subsubsection{Sampling of event type}
Using the discrete MC sampling technique, the final states for exclusive modes are sampled according to the ratios of their cross sections ($\sigma_m$) to the total cross section ($\sigma^\textrm{tot}$), i.e., $$c_m=\sigma_m(M_\textrm{hadrons})/\sigma^\textrm{tot}(M_\textrm{hadrons}),$$ where $m$ is an index for exclusive precess, and events for the remainder part, $1-\sum_mc_m$, are generated with the \lunda model.

\section{Optimization of \lunda parameters}
\subsection{Strategy to optimize the \lunda parameters}
The \lunda model parameters are optimized with the parameterized response function method. The optimal values are obtained by simultaneously fitting this function to data distributions. The idea for this method is borrowed from that implemented in the event generator tuning tool Professor and Rivet \cite{lhctune} system, which was introduced by TASSO, and later used by ALEPH, DELPHI \cite{tasso1,tasso2,aleph1,aleph2,hamacher,delphi2}, and recently by the LHC \cite{lhctune}. This method has the advantage of eliminating the problem from the so-called manual and brute-force tunings, such as the slow tuning procedure and the sub-optimal results.

An ensemble of MC samples was produced within the framework of the BesEvtGen \cite{besevtgen} event generator, and then it is subject to detector simulation with BOSS software \cite{boss}. 91 independent MC samples were prepared, each one generated with a different set of \lunda  parameters, which were randomly chosen in the parameter space around a given central point ${\bf p}_0$. All MC samples were produced with equal statistics, and were large enough so that the overall statistical uncertainties are negligible.

By including the correlations among the model parameters, the dependence of physical observable is expanded up to the quadratic term as done in Ref. \cite{delphi}, and the response function reads
\begin{eqnarray}\label{masterEq}
f({\bf p}_0+\delta {\bf p},x)=a_0^{(0)}(x)+\sum_{i=1}^{n}a_i^{(1)}(x)\delta p_i\nonumber\\
+\sum_{i=1}^{n}\sum_{j=i}^{n}a_{ij}^{(2)}(x)\delta p_i\delta p_j\approx MC({\bf p}_0+\delta {\bf p},x),
\end{eqnarray}
where $n$ is the number of parameters to be fitted, and $MC({\bf p}_0+\delta {\bf p},x)$ denotes the distribution of physical observable $x$ predicted for a given set of parameter values ${\bf p}_0+\delta {\bf p}$, where ${\bf p}_0$ is the central value and $\delta p_i$ is the deviation of the $i$-th parameter. The quadratic term in the expansion accounts for the possible correlations between the model parameters. The number of coefficients $a^{(0,1,2)}$, $L$, in the expansion is calculated with
\begin{equation}\label{nsets}
L=1+n+n(n+1)/2,
\end{equation}
and the coefficients are determined by fitting Eq. (\ref{masterEq}) to the $L$ reference simulation distributions. This fit is equivalent to solving a system of linear equations of Eq. (\ref{masterEq}). Then the optimal values of the parameters $p_i$, their errors $\sigma_i$, and their correlation coefficients $\rho_{ij}$ will be determined with a standard $\chi^2$ fit to data using package {\small\sc MINUIT} \cite{minuit}. The fit is done simultaneously for all distributions and for all bins.

To minimize statistical uncertainties, the model parameters should be fitted to the distributions that show strong dependence on the parameters under consideration and least dependence on the others. For each distribution, a quality to measure the sensitivity to the model $i$-th parameter is calculated, i.e.
\begin{equation}
S_i(x)={\delta MC(x)\over MC(x)}\Big|_{p_i}\Big/{\delta p_i\over p_i}\approx {\partial \ln MC(x)\over \partial \ln |p_i|}\Big|_{p_i},
\end{equation}
where $\delta MC(x)$ is the change of the distribution $MC(x)$ when the model parameter $p_i$ is changed by $\delta p_i$ from its central value. Sensitivity values for charged track distributions and event shapes vary within the range from -0.3 to 0.3, but the polar angle and azimuthal distributions for charged tracks are not sensitive to the change of model parameters. This is because the inclusive charged tracks are distributed isotropically over the whole phase space. Taking the sensitivity into consideration, only 12 observable distributions are kept for the model parameter fit. They are the number of photons ($N_\upgamma$), the number of charged tracks ($N_{\rm track}$), momentum of tracks ($P_{\rm track}$), $x_f=2P_z/W,~x_{\perp}=2P_{\perp}/2W$, sphericity, aplanarity, thrust, oblateness, and Fox-Wolfram moments ($H_{20},H_{30},H_{40}$) \cite{pythia6.4}, where $W$ is the total reconstructed energy of an event, and $P_{\perp}$ is the transverse momentum.

We have 12 parameters to be optimized. According to Eq. (\ref{nsets}), there are 91 coefficients, $a^{(0,1,2)}$ in Eq. (\ref{masterEq}) to be determined. Hence we need at least 91 MC samples to determine these coefficients. These were prepared with 0.5 million events for each sample. Then the dependence of response function on model parameters is established, and this analytical expression is used to simultaneously fit to the data distributions after QED background events are subtracted. In the optimization procedure, the $\chi^2$ function is defined over each bin, ie., $\chi^2\to\chi^2/N$, where $\chi^2$ values are calculated over nonempty $N$ bins. To consider the requirement of fit goodness on the multiplicity of charged tracks, this distribution is weighted with a factor of 10, while other distributions are weighted with a unitary factor. This weighted factor is chosen by requiring that the fit quality of all distributions are satisfactory.

\subsection{Event selection and fit results}
We use the data taken at $\sqrt s=$3.65 GeV to optimize the parameters. To validate these parameters, we check whether it is suitable for describing the data distribution in the energy region 2.0 -- 4.26 GeV. The QED backgrounds, e.g. $\ee\to\ee,~\gg,~\upgamma^*\upgamma^*,~\mm$, and $\uptau^+\uptau^-$ are subtracted using MC samples, and they are normalized according to their cross sections to the luminosity of data sets. The event selection criteria for light hadrons are similar to those applied to the $R$ value measurements \cite{bes2R,besrvalue}.

The selected candidates are characterized by the distributions of charged track multiplicity ($N_\textrm{track}$), track energy ($E_\textrm{track}$) and momentum ($p_\textrm{track}$), polar angle ($\cos\theta$), azimuthal angle ($\phi$), rapidity, peseudorapidity, and a set of event shapes. These distributions are normalized to one and the errors are scaled for all bins.

To consider the possible correlations between these observable quantities, different observable combinations were tried. In each combination, track observables, $N_\upgamma$, $N_{\rm track}$, $E_{\rm track}$, $x_f$ and $x_{\perp}$, must be included, while the $p_{\rm track}$ distribution or event shapes are partly included in the simultaneous fit. Generally speaking, the more observable distributions are involved in the fit, the worse fit quality one gets. To validate the resulted parameters, they are reused to generate MC samples, and compared to data.

The covariant matrix in fitting was checked, and it shows that there are strong correlations among these parameters. This indicates that the model parameters in question are not independent, which leads to some technical issues. One is the instability of the fitted values. If initial values are changed, then the fit gives a different set of parameters with almost the same fit quality. The dependence on the initial values brings about the so called multi-solution. Fortunately, we find that the produced MC distributions with these multi-solution values are in good agreement with data distributions. The correlation between the parameters means that the fitted value may be unphysical. One recipe to tackle this issue is to fix correlated parameters to the physical values, thus the fit can yield physical values for uncorrelated model parameters.
\begin{table}
\begin{center}
\caption{Optimized parameters at $\sqrt s=$3.65 GeV. The statistical errors are negligible.  $^{(2S+1)}P_J$ denotes a meson has spin $S$, orbital angular momentum $(L)$ and total spin $J$.\label{pars}}
\begin{tabular}{l c l}
\hline\hline
Parameters &  Tuned &Description\\\hline
PARJ(1)  &0.065 &Suppression of diquark-antidiquark pair production \\\hline
PARJ(2)  &0.260 &Suppression of $\mathrm{s}$ quark pair production\\\hline
PARJ(11) &0.612 &Probability that a light meson has spin 1\\\hline
PARJ(12) &0.000 &Probability that a strange meson has spin 1\\\hline
PARJ(14) &0.244 &Probability for a $^1P_1$ meson production\\\hline
PARJ(15) &0.000 &Probability for a $^3P_0$ meson production\\\hline
PARJ(16) &0.437 &Probability for a $^3P_1$ meson production\\\hline
PARJ(17) &0.531 &Probability for a $^3P_2$ meson production\\\hline
PARJ(21) &0.066 &Width of Gaussian for transverse momentum\\\hline
RALPA(15)&0.577&\lunda model parameter \\\hline
RALPA(16)&0.364&\lunda model parameter\\\hline
RALPA(17)&0.000&\lunda model parameter\\\hline\hline
\end{tabular}
\end{center}
\end{table}

\section{Validation of tuned parameters}
In the simultaneous fit to data, we have tried various combination of data distributions, which results in a few sets of parameters. To select the most optimal values, we compare the data to the MC distributions, which are generated with optimized parameters for all sets. We require that the parameters can produce MC distributions having the best fit goodness quality $\chi^2/N$, where $N$ is the total number of  bins for calculating the $\chi^2$ values. The optimal values are given in Table \ref{pars}. These values are only responsible for unknown processes other than the exclusive modes. For example, the parameter PARJ(15)=0 implies that exclusive modes have produce sufficient scalar mesons, so the \lunda model forbids the scalar meson production.

\begin{figure}[htbp]
\begin{center}
\epsfysize=10cm \epsffile{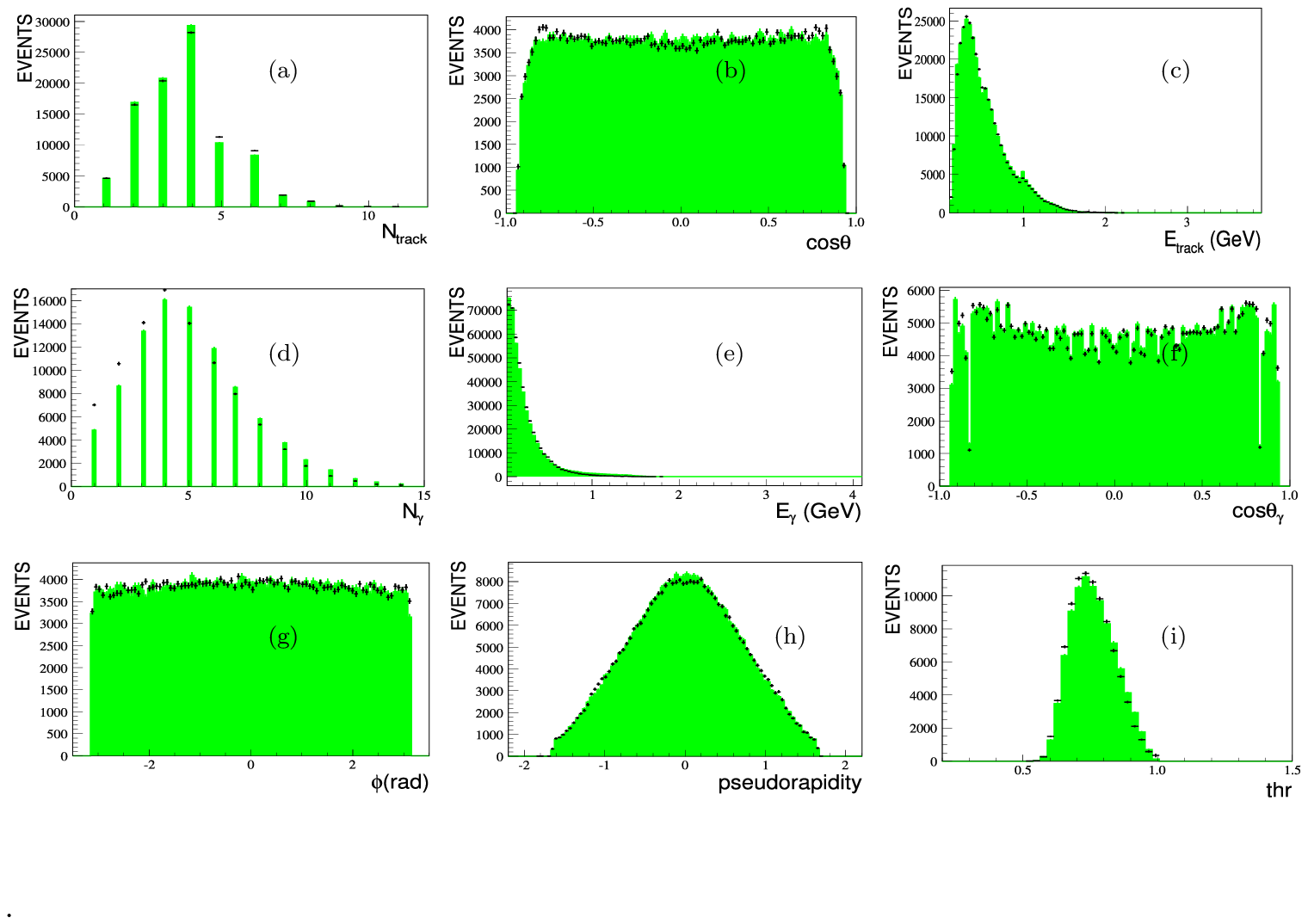}
\caption{(color online) Comparison of data to MC distributions at 3.65 GeV, where the MC sample is produced with the optimized parameters: (a) multiplicity of charged tracks, (b) cosine of polar angle of charged tracks, (c) energy of charged tracks, (d) multiplicity of photon, (e) energy of photon, (f) cosine of polar angle of photons, (g) azimuthal distribution, (h) pseudorapidity and (i) thrust. \logo \label{tuned@3.65} }
\end{center}
\end{figure}

\begin{figure}[htbp]
\begin{center}
\epsfysize=10cm \epsffile{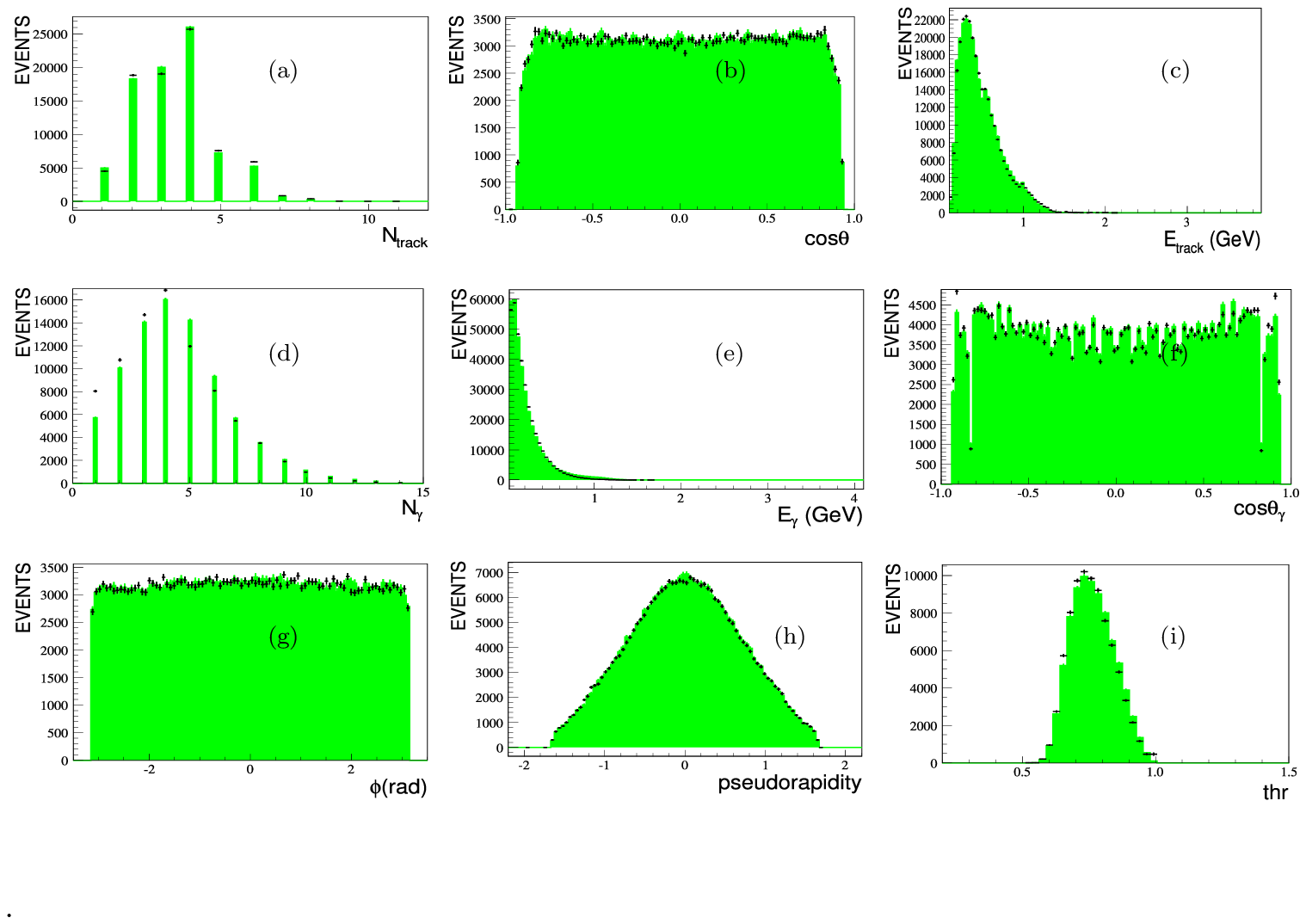}
\caption{(color online) Comparison of data to the MC distributions at 3.06 GeV, where the MC sample is produced with the optimized parameters: (a) multiplicity of charged tracks, (b) cosine of polar angle of charged tracks, (c) energy of charged tracks, (d) multiplicity of photon, (e) energy of photon, (f) cosine of polar angle of photons, (g) azimuthal distribution, (h) pseudorapidity and (i) thrust. \logo  \label{tuned@3.65ck3060}}
\end{center}
\end{figure}

Figure \ref{tuned@3.65} shows a comparison of data and MC distributions at $\sqrt s=3.65$ GeV, where the MC sample is produced with the optimized parameters. The agreement between them is satisfactory. To demonstrate the flexibility of these parameters at low energy points, we generate MC at 3.06 GeV with the same parameters, and Fig. \ref{tuned@3.65ck3060} shows comparisons between the data and MC simulation. The agreement between the data and MC distributions is acceptable. However, above the $\mathrm{D\bar D}$ threshold, we check these parameters with the data taken at 4.26, 4.23 and 4.6 GeV, and we find that the agreement between data and MC gets worse. This suggests that the optimized parameters are acceptable only below the $\mathrm{D\bar D}$ threshold. To optimize parameters above the $\mathrm{D\bar D}$ threshold, the charm meson decays will have to be added.

\begin{figure}[htbp]
\begin{center}
\epsfysize=18cm \epsffile{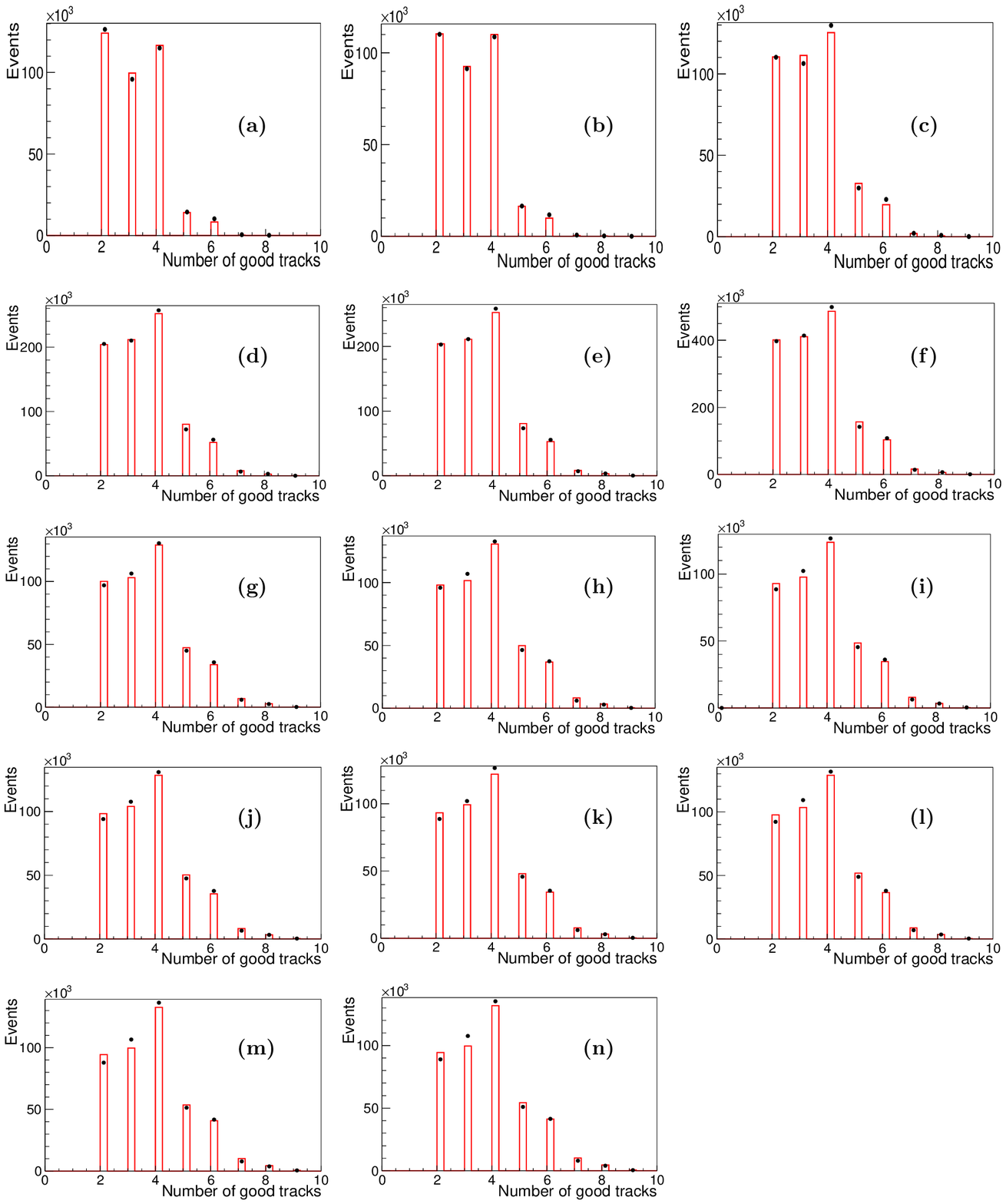}
\caption{(color online) Comparison of distributions between data and MC for the number of charged tracks at (a) 2.2324 GeV, (b) 2.4000 GeV, (c) 2.8000 GeV, (d) 3.0500 GeV, (e) 3.0600 GeV, (f) 3.0800 GeV, (g) 3.400 GeV, (h) 3.500 GeV, (i) 3.5424 GeV, (j) 3.5538 GeV, (k) 3.5611 GeV, (l) 3.6002 GeV, (m) 3.6500 GeV, (n) 3.6710 GeV. The dots denote data, and the open bars denote MC. \label{tuned@3.65cklowenergy}}
\end{center}
\end{figure}

To validate this set of parameters for the MC generation below the $\mathrm{D\bar D}$ threshold, we compare the charged track multiplicity distributions at 14 energy points from $\sqrt s=2.2324$ to $3.671$ GeV, as shown in Fig. \ref{tuned@3.65cklowenergy}.
When extending this set of parameters from 3.65 GeV to low energy points, the agreement between the data and MC multiplicity distributions gets better. This is due to the fact that the total cross section equals the sum of the exclusive ones when approaching the energy 2.0 GeV, as shown in Fig. \ref{xssum}.

\section{Discussion and summary}

To summarize, we have developed an event generator for $R$ measurement at energy scan experiments, incorporating the initial state radiation effects up to the second order correction. In the event generator, the ISR correction factor is calculated using the totally hadronic Born cross sections measured in experiments. The measured exclusive processes are generated according to their cross sections, while unknown processes are generated using the \lunda model, whose parameters are tuned with the data collected at 3.65 GeV. To validate the optimized parameters, we compare various distributions using data sets covering from energy $\sqrt s=2.2324$ to $3.671$ GeV. We conclude that the optimized parameters are valid for MC generation below the $\mathrm{D\bar D}$ threshold. Above the $\mathrm{D\bar D}$ threshold, the parameters should be optimized with the charm meson decays.

\vspace{1cm}
{\it We are grateful to Prof. Yuan Changzheng, Prof. Li Haibo, Dr. Zhu Kai and Dr. Wang Yaqian for valuable suggestions on the text revision}

.


\end{CJK*}
\end{document}